\documentclass[preprint,3p,11pt,times]{elsarticle}

\usepackage{ifpdf}
\ifpdf
  \usepackage[1.7]{bxpdfver}
\fi
\usepackage{hyperref}
\usepackage{amssymb}
\usepackage{amsmath}
\usepackage{amsfonts}
\usepackage{multirow}
\usepackage{algpseudocode}
\usepackage{xcolor}
\usepackage{orcidlink}

\journal{arXiv}

\begin{document}

\begin{frontmatter}

\title{Magnetic Resonance Simulation of Effective Transverse Relaxation ($T_2^*$)}
\author[First]{Hidenori Takeshima \orcidlink{0000-0002-2557-8533}}

\address[First]{MRI Systems Development Department, MRI Systems Division,\\
                Canon Medical Systems Corporation, Kanagawa, Japan}

\begin{abstract}

Purpose:
To simulate effective transverse relaxation ($T_2^*$) as a part of MR simulation.
$T_2^*$ consists of reversible ($T_2^{\prime}$) and irreversible ($T_2$) components.
Whereas simulations of $T_2$ are easy,
$T_2^{\prime}$ is not easily simulated if only magnetizations of individual isochromats are simulated.

Theory and Methods:
Efficient methods for simulating $T_2^{\prime}$ were proposed.
To approximate the Lorentzian function of $T_2^{\prime}$ realistically,
conventional simulators require 100+ isochromats.
This approximation can be avoided by
utilizing a linear phase model for simulating an entire Lorentzian function directly.
To represent the linear phase model,
the partial derivatives of the magnetizations with respect to the frequency axis were also simulated.
To accelerate the simulations with these partial derivatives,
the proposed methods introduced two techniques: analytic solutions, and combined transitions.

For understanding the fundamental mechanism of the proposed method,
a simple one-isochromat simulation was performed.
For evaluating realistic cases,
several pulse sequences were simulated using two phantoms with and without $T_2^{\prime}$ simulations.

Results:
The one-isochromat simulation demonstrated that $T_2^{\prime}$ simulations were possible.
In the realistic cases, $T_2^{\prime}$ was recovered as expected
without using 100+ isochromats for each point.
The computational times with $T_2^{\prime}$ simulations were only 2.0 to 2.7 times
longer than those without $T_2^{\prime}$ simulations.
When the above-mentioned two techniques were utilized,
the analytic solutions accelerated 19 times, and
the combined transitions accelerated up to 17 times.

Conclusion:
Both theory and results showed that
the proposed methods simulated $T_2^{\prime}$ efficiently by utilizing
a linear model with a Lorentzian function,
analytic solutions, and combined transitions.

\end{abstract}


\end{frontmatter}

\section{Introduction}

Acquisition processes of an MR system
can be simulated with the Bloch equations and their extensions\cite{Bloch,Torrey,McConnell}.
Software for simulating these processes is often referred to as MR simulators.
The simulators are useful for developing, prototyping, and evaluating
various components of MR systems such as pulse sequences, image reconstruction, and postprocessing.
Pulse sequences are simulated by updating
time-dependent magnetizations of isochromats,
and integrating these magnetizations whenever data sampling is requested.
The output samples are integrals of these magnetizations in spatial and frequency axes.

There are two representations of isochromats: discrete and continuous representations.
The simulators using the discrete representations%
\cite{Taniguchi1,Taniguchi2,SIMRI,JEMRIS,MRISIMUL,MRiLab,Kose,coreMRI,Scholand,KomaMRI,Takeshima,Takeshima2}
assigned discrete points in continuous space to individual isochromats.
In these simulators, the integrals were implemented as summations.
In contrast, the simulators using the continuous representation%
\cite{Jochimsen}
assigned small regions in continuous space to individual isochromats.
All points in each region were treated as continuous.
The magnetizations of these continuous points were represented as a constant magnitude with a linear phase model.

The purpose of this paper is to develop a precise and efficient simulation method of
the effective transverse relaxation ($T_2^*$).
$T_2^*$ consists of irreversible ($T_2$) and reversible ($T_2^{\prime}$) components.
Whereas it is easy to simulate $T_2$,
there were no known ways to simulate $T_2^{\prime}$ efficiently.
Since many simulators used the discrete representations,
a common way for simulating $T_2^{\prime}$ was
to add a Lorentzian noise to magnetic field inhomogeneity\cite{JEMRIS,MRiLab,Takeshima}.
When the simulated number of discrete isochromats was increased,
its behaviors approached asymptotically to the true behaviors.
However, as explained later, realistic $T_2^{\prime}$ simulations require 100+ isochromats for each point.
In practice, the number of isochromats cannot be increased sufficiently
since computational costs are roughly proportional to the number of isochromats.
Even without increasing the number, many MR simulators 
needed to reduce computational costs and thus relied on hardware acceleration
such as multi-threading \cite{JEMRIS,Kose,Takeshima,Takeshima2},
single instruction multiple data (SIMD) \cite{Kose,Takeshima,Takeshima2},
general-purpose graphics processing unit (GPGPU) \cite{MRISIMUL,MRiLab,Kose,KomaMRI},
computer clusters \cite{SIMRI,JEMRIS},
and cloud computing \cite{coreMRI}.

In this paper,
efficient methods for simulating $T_2^{\prime}$
using a continuous representation were proposed.
To simulate $T_2^{\prime}$ efficiently, a theory for simulating $T_2^{\prime}$ with only one isochromat was derived.
For understanding the fundamental mechanism of the proposed method,
this theory was experimentally confirmed by comparing the difference
between discrete and continuous representations
using a simple case.
Realistic $T_2^*$ simulations were also demonstrated using a house-made MR simulator.

\section{Theory}

\subsection{Baseline Methods}
\begin{figure}[t]%
\centering%
\includegraphics[width=\linewidth]{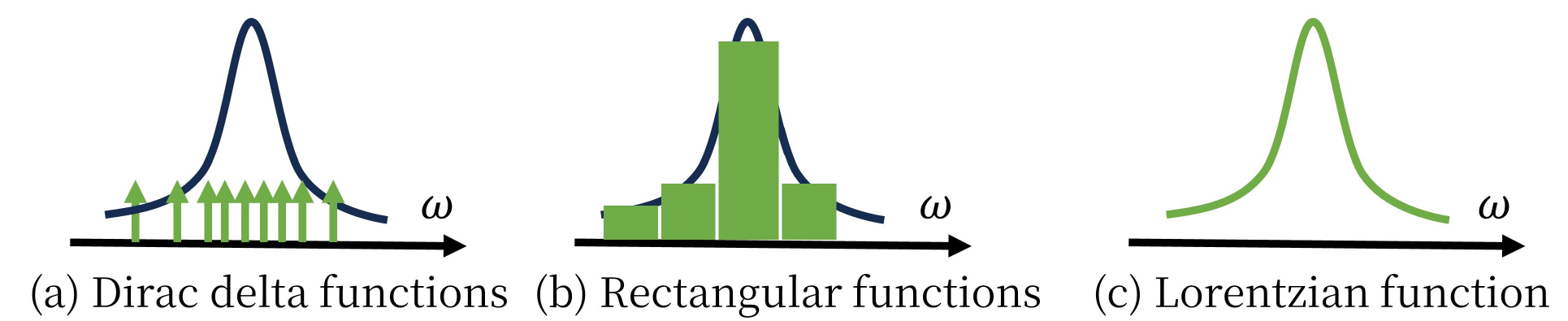}%
\caption{Various $T_2^{\prime}$ models.
(a) A set of the Dirac delta functions.
(b) A set of the rectangular functions.
(c) A Lorentzian function.
Whereas the models (a) and (b) approximates a Lorentzian function to be simulated with a set of functions,
the model (c) uses the Lorentzian function to be simulated directly.%
}\label{figLorentizian}
\end{figure}
\begin{figure}[t]%
\centering%
\includegraphics[width=\linewidth]{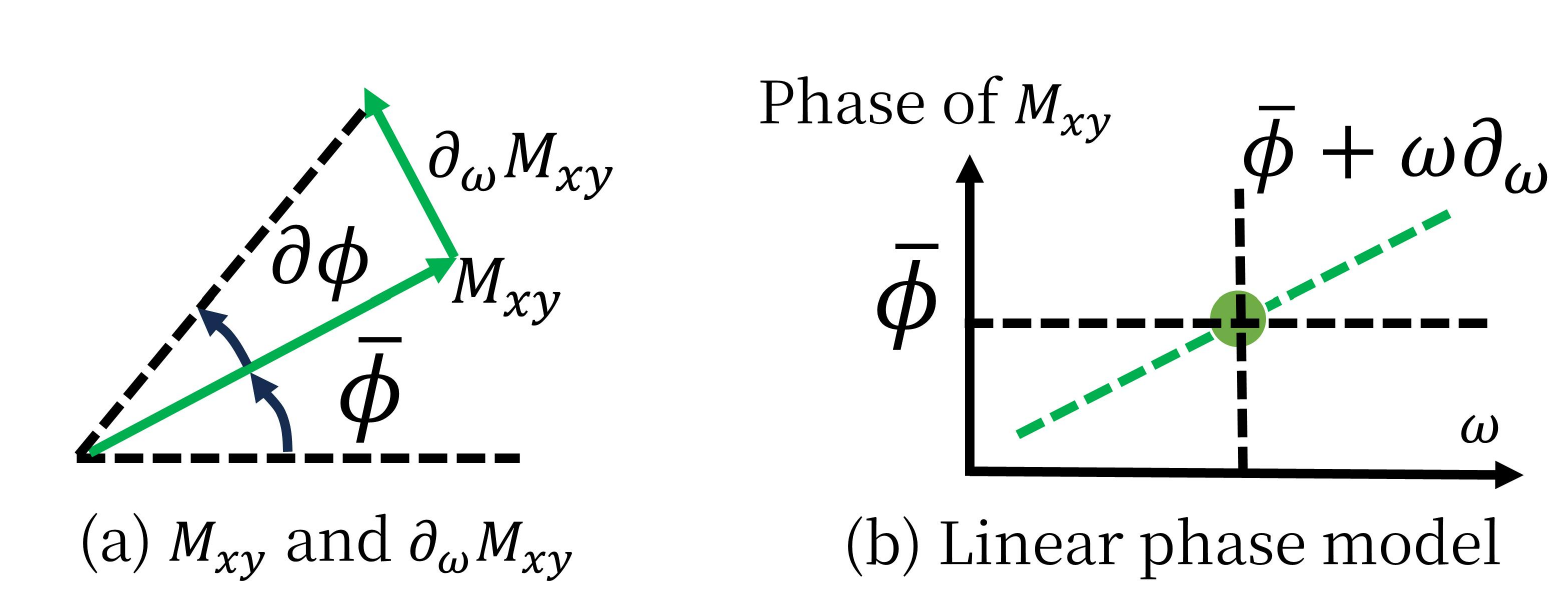}%
\caption{A continuous representation.
(a) Relationship between a transverse magnetization and
its partial derivative with respect to $\omega$.
The transverse magnetization is orthogonal to its partial derivative.
(b) A linear phase model.
The transverse magnetization $M_{xy}$ is modeled as
a constant magnitude with a phase linearly changed in the $\omega$ axis.%
}\label{figContinuous}
\end{figure}
\begin{figure}[t]%
\centering%
\includegraphics[width=7cm]{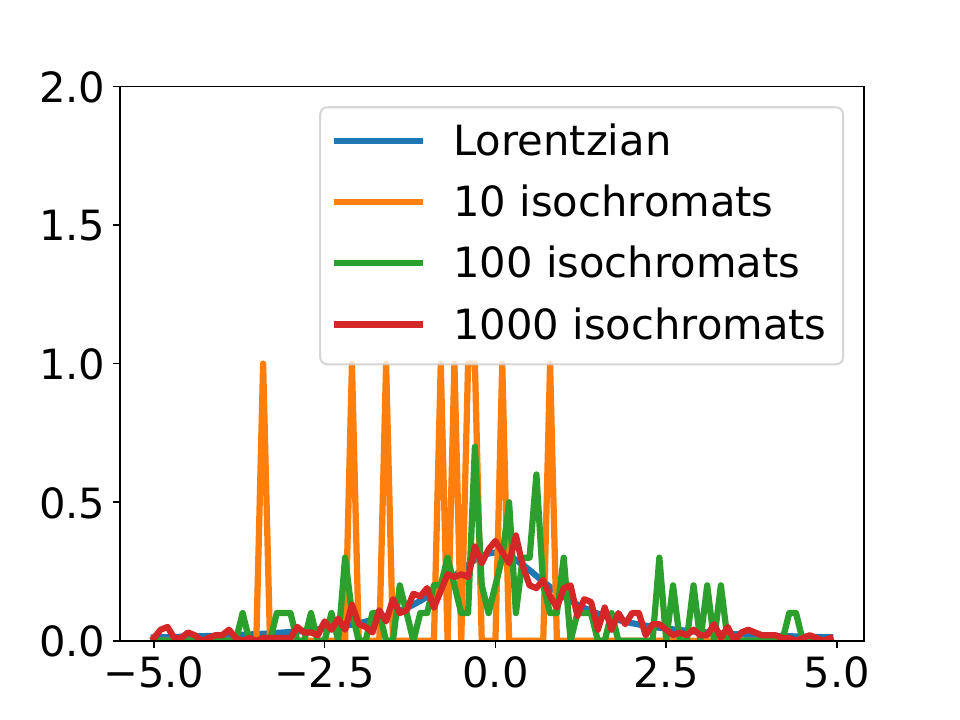}%
\caption{An example case of approximating a Lorentzian function using a set of thin rectangular functions.
The function approximated with 10 isochromats is far from the Lorentzian function.
There are still spike-like errors in the function approximated with 100 isochromats.
The function approximated with 1000 isochromats is similar to the Lorentzian function.
These functions show that at least 100+ rectangular functions are required for better approximation.%
}\label{figLorentizianSampling}
\end{figure}

Let the direction of the static magnetic field be the $z$ axis.
Let the remaining axes representing the rotating frame around the $z$ axis be the $x$ and $y$ axes.
For the $k$-th isochromat ($k=1, \ldots, K$),
the Bloch equations \cite{Bloch} were defined as
\begin{equation}\label{eq:Bloch}
\frac{d\boldsymbol{M}_3(k,t)}{dt} = \gamma (\boldsymbol{M}_3(k,t) \times \boldsymbol{B}(k,t)) +
   \mathrm{diag}(1/T_2 (k), 1/T_2 (k), 1/T_1 (k)) 
   ((0, 0, 1)^T - \boldsymbol{M}_3(k,t))
\end{equation}
where
$\gamma$,
$\boldsymbol{M}_3(k,t) = (M_x(k,t), M_y(k,t), M_z(k,t))^T$,
$\boldsymbol{B}(k,t) = (B_x(k,t), B_y(k,t), B_z(k,t))^T$,
$T_1 (k)$, and $T_2 (k)$
represent
the gyromagnetic ratio,
the \emph{normalized} time-dependent magnetization vector at the time $t$,
the magnetic field vector at the time $t$,
longitudinal relaxation time, and
transverse relaxation time,
respectively.
The relaxation times are given as a part of a phantom.
The time-dependent magnetization vector $\boldsymbol{M}_3(k,t)$ is updated in a simulation process.

The magnetic field vector $\boldsymbol{B}(k,t)$ consists of $\boldsymbol{B}_1(k,t)$ and $B_z(k,t)$.
$\boldsymbol{B}_1(k,t) = B_x(k,t) + iB_y(k,t)$ represents the radiofrequency (RF) pulses.
$B_z(k,t)$ consists of several components and is represented as
\begin{equation}\label{eq:B_z}
  \gamma B_z(k,t) = \gamma G(\boldsymbol{r}(k,t),t) + \gamma \Delta B_0(\boldsymbol{r}(k,t)) + d(k) + \omega
\end{equation}
where
$\boldsymbol{r}(k,t) = (r_x(k,t), r_y(k,t), r_z(k,t))^T$ represents the location of the isochromat,
$G(\boldsymbol{r}(k,t),t)$ represents the magnetic fields from the gradient coils,
$\Delta B_0(\boldsymbol{r}(k,t))$ represents the magnetic field inhomogeneity,
$d(k)$ represents the chemical shift, and
$\omega$ represents the perturbation come from the reversible transverse relaxation. 
In the cases of static isochromats, the time $t$ of $\boldsymbol{r}(k,t)$ can be omitted.
In the cases of linear gradient coils with no gradient imperfections,
$G(\boldsymbol{r}(k,t),t)$ is defined as
\begin{equation}\label{eq:G_coils}
G(\boldsymbol{r}(k,t),t) = G_x(t) r_x(k,t) + G_y(t) r_y(k,t) + G_z(t) r_z(k,t)
\end{equation}
where $G_x(t)$, $G_y(t)$, and $G_z(t)$ represent gradients of a pulse sequence along $x$, $y$, and $z$ axes, respectively.

When RF pulses $\boldsymbol{B}_1(k,t)$ are not used (referred to as without-RF subsequences),
the magnetization at $t$ can be computed analytically as
\begin{align}
\boldsymbol{M}_{xy} (k,t) &= \boldsymbol{M}_{xy} (k,t_0) \exp \Bigg(-\frac{t-t_0}{T_2(k)} \Bigg) \exp \Bigg( \int_{t_0}^t -i\gamma B_z(k,t)dt \Bigg) \label{eq:analytic_xy}\text{, and} \\
   M_z (k,t) &= 1+(M_z(k,t_0) - 1) \exp \Bigg(-\frac{t-t_0}{T_1(k)} \Bigg) \label{eq:analytic_z}
\end{align}
where $\boldsymbol{M}_{xy} (k,t)$ is defined as $\boldsymbol{M}_{xy} (k,t) = M_x(k,t) + i M_y (k,t)$.

When RF pulses $\boldsymbol{B}_1(k,t)$ are used (referred to as with-RF subsequences),
iterative update is required for computing the time-dependent magnetization vector.
If the magnetic field vector $B(k,t)$ is piecewise constant between $t$ and $t+\Delta t$,
the transition of the time-dependent magnetization vector
between $t$ and $t+\Delta t$ can be represented as matrix multiplications by introducing
the 4-dimensional vector $\boldsymbol{M}_4(k,t) = (M_x(k,t), M_y(k,t), M_z(k,t), 1)^T$.
For a small duration $\Delta t$, the transition
from $t$ to $t+\Delta t$ can be written as
\begin{equation}\label{eq:update}
\boldsymbol{M}_4(k,t+\Delta t) = \boldsymbol{D}_4(k, \Delta t) \boldsymbol{R}_4(k, t, t+\Delta t)\boldsymbol{M}_4(k,t)
\end{equation}
where $\boldsymbol{D}_4(k, \Delta t)$ and $\boldsymbol{R}_4(k, t, t+\Delta t)$ are $4 \times 4$ matrices given in \ref{appendix:transition}.

The computation of Eq. (\ref{eq:update}) can be accelerated by
computing combined transitions once and
using the combined transitions multiple times%
\cite{Taniguchi1,Taniguchi2,Scholand,Takeshima}
since many pulse sequences use same RF pulses repeatedly.
By combining $\boldsymbol{D}_4(k, \Delta t) \boldsymbol{R}_4(k, t, t+\Delta t)$ at $t=t_0,t_1,\ldots,t_{N_{RF}-1}$,
a combined transition from $t_0$ to $t_{N_{RF}}$ can be represented as
\begin{equation}\label{eq:combined_update}
  \boldsymbol{M}_4(k,t_{N_{RF}}) =
    \boldsymbol{C}_4(k,t_0,t_{N_{RF}})
    \boldsymbol{M}_4(k,t_0)
\end{equation}
where
\begin{equation}\label{eq:combined_transition}
\boldsymbol{C}_4(k,t_0,t_{N_{RF}}) =
\boldsymbol{D}_4(k, \Delta t) \boldsymbol{R}_4(k, t_{N_{RF}-1}, t_{N_{RF}})
\cdots
\boldsymbol{D}_4(k, \Delta t) \boldsymbol{R}_4(k, t_0, t_1)
\end{equation}
represents the combined transition from $t_0$ to $t_{N_{RF}}$.

When a pulse sequence indicates use of analog-to-digital converters (ADCs),
integrals of transverse magnetizations at specified times are computed.
For the $l$-th receiver coil, the integral $A(l,t)$ can be represented as
\begin{equation}\label{eq:adc}
A(l,t) = \sum_{k=1}^{K} \int_{-\infty}^{\infty} \int_{-\infty}^{\infty} \int_{-\infty}^{\infty} \int_{-\infty}^{\infty}
   w_l(\boldsymbol{r}(k,t), t) M_0(k) \boldsymbol{M}_{xy}(k,t)
   f_x(x - r_x) f_y(y - r_y) f_z(z - r_z) f_{\omega}(\omega) d\omega dz dy dx \text{.}
\end{equation}
where $w_l(\boldsymbol{r}(k,t), t)$ represents the receiver coil sensitivity at the location $\boldsymbol{r}(k,t)$,
$M_0(k)$ represents the scaling factor of the $k$-th isochromat, and
$f_p(p)$ represents the magnitude and phase function relative to the rectangular center
along the $p \in {x,y,z,\omega}$ axis.
In the discrete representations shown in Fig. \ref{figLorentizian}(a),
$f_p(p)$ are
the Dirac delta functions for all axes.
In these cases, Eq. (\ref{eq:adc}) is simplified as
\begin{equation}\label{eq:adc_simplified}
A(l,t) = \sum_{k=1}^{K} w_l(\boldsymbol{r}(k,t), t) M_0(k) \boldsymbol{M}_{xy}(k,t)\text{.}
\end{equation}

To use continuous representations,
Jochimsen et al.\cite{Jochimsen} used a linear-phase model
for representing $f_p(p)$ as a parametric function (Fig. \ref{figContinuous}).
For a region covered by an isochromat, its magnitudes are modeled as constant.
The phases $\phi_p$ are modeled as
$\phi_p(p) = \overline{\phi_p} + p \partial_p \phi_p$
where $\overline{\phi_p}$ represents the phase at the center, and
$\partial_p$ represents a partial derivative with respect to $p$.
The linear coefficient $\partial_p \phi_p$ can be computed as
\begin{equation}\label{eq:T2prime_phi}
\partial_p \phi_p = \frac{M_x(k,t) \partial_p M_y(k,t) - M_y(k,t) \partial_p M_x(k,t)}{ \sqrt{M_x(k,t)^2 + M_y(k,t)^2} } \text{.}
\end{equation}

To compute the linear coefficient,
it is necessary to simulate $\partial_p \boldsymbol{M}_3(k,t)$ for computing $\partial_p \phi_p$.
Jochimsen et al.\cite{Jochimsen} simulated $\partial_p \boldsymbol{M}_3(k,t)$
by approximating $\boldsymbol{B}_1(k,t)$ as piecewise constants for individual isochromats,
and computing approximated partial derivatives of the transition Eq. (\ref{eq:update}).
The partial derivatives of the transition from $t$ to $t+\Delta t$ can be written as
\begin{equation}\label{eq:update_derivatives}
\partial_p \boldsymbol{M}_3(k,t+\Delta t)
=
\boldsymbol{D}_3(k, \Delta t)
\boldsymbol{R}_{3,xy}(k, t, t+\Delta t)
\{
\partial_p \boldsymbol{R}_{3,z}(k, t, t+\Delta t)
\boldsymbol{M}_3(k,t)
+
\boldsymbol{R}_{3,z}(k, t, t+\Delta t)
\partial_p \boldsymbol{M}_3(k,t)
\}
\text{.}
\end{equation}
The approximation process and matrices used in Eq. (\ref{eq:update_derivatives}) are given in \ref{appendix:transition}.
In Eq. (\ref{eq:update_derivatives}),
the last element of $\partial_p \boldsymbol{M}_4(k,t)$ is omitted since it is 0 for all $p$'s.

Jochimsen et al. treated isochromats as rectangular functions in both spatial and frequency axes.
In their model, $f_p(p)$ is defined as
$(1/L) \exp (i \phi_p(p))$ if $|p| < L/2$ and 0 otherwise (Fig. \ref{figLorentizian}(b)).

\subsection{Continuous Representation for Simulating $T_2^{\prime}$}\label{subsec:continuous}

Conventional MR simulators
sampled $\omega$ from
the Lorentzian whose half-width at half-minimum was $1/T_2^{\prime}$.
In these simulators,
100+ isochromats per a discrete point
are required in the $\omega$ axis for simulating $T_2^{\prime}$ realistically
as shown in Fig. \ref{figLorentizianSampling}.
Use of the continuous representation with rectangular functions cannot solve this problem.

Rather than using a set of points or rectangular functions for representing $T_2^{\prime}$,
$f_{\omega}(\omega)$ is modeled as
\begin{equation}\label{eq:T2prime_g}
  f_{\omega}(\omega) = \ell(\omega; 0, 1/T_2^{\prime}(k)) \exp(i \omega \partial \phi_{\omega}(\omega) )
\end{equation}
where $\ell(\omega; 0, 1/T_2^{\prime}(k))$ represents a Lorentzian function whose mode and scale parameter are 0 and $1/T_2^{\prime}(k)$, respectively
(Fig. \ref{figLorentizian}(c)).
The remaining functions $g_x(x)$, $g_y(y)$, and $g_z(z)$ are the Dirac delta function.

By computing the integration parts of Eq. (\ref{eq:adc}) with Fourier transforms,
the integration value $A(l,t)$ becomes
\begin{equation}\label{eq:adc_proposed}
  A(l,t) = \sum_{k=1}^{K}
   w_l(\boldsymbol{r}(k,t), t) M_0(k) \boldsymbol{M}_{xy}(k,t)
   \exp \Big(- (1/T_2^{\prime}(k)) |\partial \phi_{\omega}(\omega)| \Big) \text{.}
\end{equation}
Pseudo-code for simulation with Eq. (\ref{eq:adc_proposed}) is given in Supplementary Material S1.

\subsection{Analytic Solutions of Without-RF Subsequences for Computing Partial Derivatives}\label{subsec:analytic}

In the cases of without-RF subsequences,
the partial derivatives of Eqs. (\ref{eq:analytic_xy}) and (\ref{eq:analytic_z}) can be computed analytically for accelerating simulations.
The partial derivative of Eq. (\ref{eq:analytic_xy}) with respect to $p$ can be computed as
\footnotesize
\begin{equation}\label{eq:analytic_xy_derivative_full}
   \partial_p \boldsymbol{M}_{xy} (k,t) =
      \exp \Bigg(-\frac{t-t_0}{T_2(k)} \Bigg)
      \Bigg(
      (\partial_p \boldsymbol{M}_{xy} (k,t_0))  \exp \Bigg( \int_{t_0}^t -i\gamma B_z(k,t)dt \Bigg)
      +
      \boldsymbol{M}_{xy} (k,t_0)  \Bigg\{ \partial_p \exp \Bigg( \int_{t_0}^t -i\gamma B_z(k,t)dt \Bigg) \Bigg\}
      \Bigg)
   \text{.}
\end{equation}
\normalsize
In the case of $p = \omega$, Eq. (\ref{eq:analytic_xy_derivative_full}) can be simplified as
\footnotesize
\begin{equation}\label{eq:analytic_xy_derivative}
   \partial_{\omega} \boldsymbol{M}_{xy} (k,t) =
      \exp \Bigg(-\frac{t-t_0}{T_2(k)} \Bigg)
      \Bigg(
      (\partial_{\omega} \boldsymbol{M}_{xy} (k,t_0))  \exp \Bigg( \int_{t_0}^t -i\gamma B_z(k,t)dt \Bigg)
      - i (t - t_0)
      \boldsymbol{M}_{xy} (k,t_0) \exp \Bigg( \int_{t_0}^t -i\gamma B_z(k,t)dt \Bigg)
      \Bigg)
   \text{.}
\end{equation}
\normalsize
since $\partial_{\omega} \int_{t_0}^t \gamma B_z(k,t) dt$ is $(t - t_0)$ according to Eq. (\ref{eq:B_z}).

The partial derivative of Eq. (\ref{eq:analytic_z}) with respect to $p$ can be computed as
\begin{equation}\label{eq:analytic_z_derivative}
   \partial_p M_z(k,t) = \partial_p M_z(k,t_0) \exp \Bigg(-\frac{t-t_0}{T_1(k)} \Bigg)
   \text{.}
\end{equation}
Pseudo-code for simulation with Eq. (\ref{eq:analytic_xy_derivative}) and Eq. (\ref{eq:analytic_z_derivative}) is given in Supplementary Material S1.

\subsection{Combined Transitions for Computing Partial Derivatives}\label{subsec:combined}

In the cases of with-RF subsequences,
it is useful to compute combined transitions and apply them repeatedly for accelerating simulations.
To rewrite Eqs. (\ref{eq:update}) and (\ref{eq:update_derivatives}) for systematic matrix operations,
non-zero elements of both $\boldsymbol{M}_4(k,t)$ and $\partial_{p} \boldsymbol{M}_3(k,t)$
are merged as a 7-dimensional vector
\begin{equation}\label{eq:vector7}
  \boldsymbol{M}_7(k,t) = (M_x(k,t), M_y(k,t), M_z(k,t), 1, \partial_{p} M_x(k,t), \partial_{p}M_y(k,t), \partial_{p}M_z(k,t))^T
  \text{.}
\end{equation}
By using this vector, Eq. (\ref{eq:update_derivatives}) can be represented as
\begin{equation}\label{eq:update_derivatives_vector7}
\boldsymbol{M}_7(k,t+\Delta t) =
  \boldsymbol{D}_7(k, \Delta t)
  \boldsymbol{R}_7(k, t, t+\Delta t)
  \boldsymbol{M}_7(k, t)
  \text{.}
\end{equation}
The $7 \times 7$ matrices introduced in Eq. (\ref{eq:update_derivatives_vector7}) are given in \ref{appendix:transition}.
Since Eq. (\ref{eq:update_derivatives_vector7}) is same as Eq. (\ref{eq:update}) except for the matrix sizes,
the combined transition $\boldsymbol{C}_7(k,t_0,t_{N_{RF}})$ can be utilized for accelerating
update of $\boldsymbol{M}_7(k,t)$ with existing reuse methodology\cite{Taniguchi1,Taniguchi2,Scholand,Takeshima}.
Similar to Eq. (\ref{eq:combined_transition}), $\boldsymbol{C}_7(k,t_0,t_{N_{RF}})$ can be represented as
\begin{equation}\label{eq:combined_transition7}
\boldsymbol{C}_7(k,t_0,t_{N_{RF}}) =
\boldsymbol{D}_7(k, \Delta t) \boldsymbol{R}_7(k, t_{N_{RF}-1}, t_{N_{RF}})
\cdots
\boldsymbol{D}_7(k, \Delta t) \boldsymbol{R}_4(k, t_0, t_1)
\text{.}
\end{equation}

Since the size of matrices are extended from $4 \times 4$ to $7 \times 7$,
straightforward computation of $\boldsymbol{C}_7(k,t_0,t_{N_{RF}})$ increases the computational times.
To avoid $7 \times 7$ matrix multiplications in computing $\boldsymbol{C}_7(k,t_0,t_{N_{RF}})$,
simulations of 7 linearly independent vectors of $\boldsymbol{M}_7(k,t)$ 
with Eqs. (\ref{eq:update}) and (\ref{eq:update_derivatives_vector7}) can be used instead.
An example set of these vectors is 7 unit vectors whose fourth elements are set to 1.

Pseudo-code for efficient computation of $\boldsymbol{C}_7(k,t_0,t_{N_{RF}})$ is given in Supplementary Material S1.

\section{Methods}

\subsection{A Simple Case for Understanding the Fundamental Mechanism}

A simple case was implemented and simulated for better understanding of the proposed method.
In this case, there is only one isochromat
whose physical values were
$M_0 = 1$, $T_1 = 1/10$, $T_2 = 1/50$, and $T_2^{\prime} = 1/200$
($T_2^* = 1/250$ since $1/T_2^* = 1/T_2 + 1/T_2^{\prime}$).
The pulse sequence to be simulated was a 15-millisecond Carr-Purcell-Meiboom-Gill (CPMG) sequence with no gradients.
This sequence consisted of
the first with-RF subsequence, the first without-RF subsequence,
the second with-RF subsequence, and the second without-RF subsequence.
The first and second with-RF subsequences contained
2-millisecond sinc RF pulses whose flip angles were adjusted to 90 and 180 degrees, respectively.
The phase of the second with-RF subsequence was set to 90 degrees.
The durations of the first and second without-RF subsequences were 3 and 8 milliseconds, respectively.
There were no RF pulses in these without-RF subsequences.

In this simulation, all with-RF subsequences were simulated with
Eqs. (\ref{eq:update}) and (\ref{eq:update_derivatives_vector7}).
All without-RF subsequences were simulated with
Eqs. (\ref{eq:analytic_xy}), (\ref{eq:analytic_z}),
(\ref{eq:analytic_xy_derivative}) and (\ref{eq:analytic_z_derivative}).
Temporal dynamics of $B_x$, $B_y$, $M_x$, $M_y$, $M_z$,
$\partial_{\omega} M_x$,
$\partial_{\omega} M_y$,
$\partial_{\omega} M_z$, and
measured transverse magnetizations to be sampled were plotted.
The measured transverse magnetizations were computed for both simulations with and without $T_2^{\prime}$.

\subsection{Phantoms with Multiple Isochromats}
\begin{figure}[t]%
\centering%
\includegraphics[width=5cm]{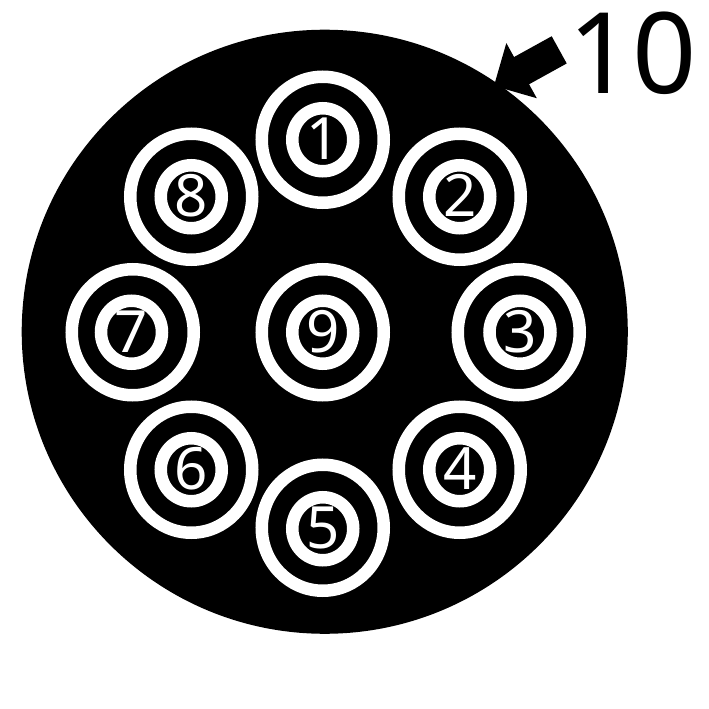}%
\caption{The layout of the circles-$T_2^{*}$ phantom.
There were a large cylinder and 9 pairs of concentric cylinders.
The $T_1$, $T_2$, and $T_2^{\prime}$ values of all cylinders are given in Table \ref{tableCircles}.%
}\label{figCircles}
\end{figure}
\begin{table}[t]%
\caption{The $T_1$, $T_2$, and $T_2^{\prime}$ values of the cylinders used in the circles-$T_2^{*}$ phantom.
All values are given in second.
Its layout is given in Fig. \ref{figCircles}.%
}\label{tableCircles}
\centering%
\begin{tabular}{|l|r|r|r|r|} \hline
Cylinder & $T_1$ & $T_2$ & $T_2^{\prime}$ (Outer) & $T_2^{\prime}$ (Inner) \\ \hline
\#1  & 0.20 & 0.05 & 0.04 & 0.02 \\ \hline
\#2  & 0.40 & 0.05 & 0.04 & 0.02 \\ \hline
\#3  & 0.60 & 0.05 & 0.04 & 0.02 \\ \hline
\#4  & 0.80 & 0.05 & 0.04 & 0.02 \\ \hline
\#5  & 1.00 & 0.05 & 0.04 & 0.02 \\ \hline
\#6  & 1.00 & 0.10 & 0.08 & 0.04 \\ \hline
\#7  & 0.80 & 0.10 & 0.08 & 0.04 \\ \hline
\#8  & 0.60 & 0.10 & 0.08 & 0.04 \\ \hline
\#9  & 0.40 & 0.10 & 0.08 & 0.04 \\ \hline
\#10 & 2.00 & 0.05 & 0.08 & 0.04 \\ \hline
\end{tabular}%
\end{table}

Two phantoms named circles-$T_2^{*}$ and brain-$T_2^{*}$ were created for
evaluating the proposed method in realistic conditions.
The numbers of isochromats were 20 and 284 million for circles-$T_2^{*}$ and brain-$T_2^{*}$, respectively.

The circles-$T_2^{*}$ phantom was a numeric phantom which consisted of
a large cylinder and 9 concentric cylinders which contained two cylinders with different diameters.
These concentric cylinders were put in the large cylinder.
All $M_0$ values in the large cylinder were shared.
In each concentric cylinder, $T_1$ and $T_2$ values of its two cylinders were shared.
$T_2^{\prime}$ values of its two cylinders were not shared.
Therefore, each concentric cylinder should be visible
as the single cylinder in the cases of spin-echo sequences, and
as the concentric cylinders in the cases of gradient-echo sequences.
All $\Delta B_0$ values were set to 0 Hz.
The matrix and spatial sizes of the phantom were
$960 \times 960 \times 48$ and
$240 \text{mm} \times 240 \text{mm} \times 12 \text{mm}$, respectively.
The spatial resolution of this phantom was $0.25 \text{mm} \times 0.25 \text{mm} \times 0.25 \text{mm}$.
The layout and parameters of this phantom are given in Fig. \ref{figCircles} and Table \ref{tableCircles}, respectively.

The brain-$T_2^{*}$ phantom was generated from multi-slice images of a brain and an artificial $T_2^{\prime}$ map.
The matrix and spatial sizes of the phantom were
$256 \times 240 \times 48$ and
$256 \text{mm} \times 240 \text{mm} \times 240 \text{mm}$, respectively.
The multi-slice images were acquired in a volunteer scan.
These images were same as the images used in a previous paper\cite{Takeshima}.
The $M_0$, $T_1$, and $T_2$ maps were generated from these images
using a parameter mapping method similar to the method.
The details of the parameter mapping method were given in Supplementary Material S2.
All $\Delta B_0$ values were set to 0 Hz.
The artificial $T_2^{\prime}$ map was created as the following two steps.
First, the map was filled with $T_2^{\prime}$ of 1/5.
Second, two dots whose sizes were $3 \times 3 \times 3$ pixels were overwritten with $T_2^{\prime}$ of 1/100.
Therefore, signals of these dots were expected to be decayed quickly in the cases of gradient-echo sequences.
In the simulation, each voxel was divided into $4 \times 4 \times 20$ subvoxels
for adjusting spatial resolution of this phantom to $0.25 \text{mm} \times 0.25 \text{mm} \times 0.25 \text{mm}$.
The volunteer scan was approved by our institutional review board and informed consent was obtained from the volunteer.

\subsection{Evaluations with Multiple Isochromats}

\begin{table}[t]%
\caption{Pulse sequences used in realistic simulations.
These pulse sequences were available as examples of the Pulseq\cite{Pulseq}.
Abbreviations: T1W, T1-weighted,
T2W, T2-weighted,
SE, spin echo,
RARE, rapid acquisition with relaxation enhancement,
EPI, echo planar imaging.}\label{tablePulseq}
\centering%
\begin{tabular}{|l|r|r|r|r|} \hline
 & SPGR & RARE & EPI & Spiral \\ \hline
Sequence name & gre.seq & tse.seq & epi.seq & spiral.seq \\ \hline
\multirow{2}{*}{RF pulse type} & \multirow{2}{*}{sinc} & \multirow{2}{*}{sinc} & \multirow{2}{*}{sinc} & Gaussian (fat sat), \\
& & & & sinc (excite) \\ \hline
Num. samples in RF & \multirow{2}{*}{3000 (excite)} & 2500 (excite) & \multirow{2}{*}{3000 (excite)} & 8000 (fat sat) \\
pulses (1 us/sample) & & 2000 (refocus) & & 3000 (excite) \\ \hline
Num. with-RF & 128 & 153 & 3 & 8 \\ \hline
Num. unique RF pulses & 1 & 2 & 3 & 5 \\ \hline
Num. repeated RF pulses & 127 & 151 & 0 & 3 \\ \hline
FOV (mm $\times$ mm) & $256 \times 256$ & $256 \times 256$ & $220 \times 220$ & $256 \times 256$ \\ \hline
Slice thickness (mm) & 3 & 5 & 3 & 3 \\ \hline
TR (ms) & 12 & 2000 & N/A & N/A \\ \hline
Approx. TE (ms) & 5 & 100 & 26 & 3 \\ \hline
Num. readout samples&128&128&64&13000 \\ \hline
\multirow{2}{*}{Num. phase encodes} & \multirow{2}{*}{128} & 128, ETL=16 & 64 & N/A \\
 & & 1 dummy shot & & \\ \hline
Num. slices&1&1&3&4 \\ \hline
Readout sampling ratio & \multirow{2}{*}{25} & \multirow{2}{*}{50} & \multirow{2}{*}{4} & \multirow{2}{*}{1.6} \\
(us/sample)& & & & \\ \hline
Reconstruction matrix& $256 \times 256$ & $256 \times 256$ & $256 \times 256$ & $256 \times 256$ \\ \hline
Oversampling factor in gridding&2.0&2.0&4.5&2.0 \\ \hline
\end{tabular}%
\end{table}

The proposed method was implemented as a part of
house-made software named a virtual MR scanner (VMRscan)\cite{Takeshima}.
In the simulations using the VMRscan,
multi-threading and SIMD instructions were used on a central processing unit (CPU).
When with-RF subsequences were simulated, the step size of the simulations was set to 1 microsecond
for compatibility with the Pulseq\cite{Pulseq}.
The brain-$T_2^{*}$ phantom was split into 8 partial phantoms for reducing memory footprint.
The simulations with individual partial phantoms were processed sequentially,
and the sampled data via simulated ADCs were merged for image reconstruction.
The computational times using the brain-$T_2^{*}$ phantom
were calculated by summing measured computational times of all partial phantoms.
The software was run on a central processing unit (CPU) with 8 performance cores,
16 efficient cores and 32 processor threads.
The frequencies of the CPU were 3.2 GHz for the performance cores and 2.4 GHz for the efficient cores.
These cores were dynamically boosted up to 6.0 GHz.

To evaluate the efficiency of the analytic solutions given as 
Eqs. (\ref{eq:analytic_xy_derivative}) and (\ref{eq:analytic_z_derivative}),
computational times of an FID sequence were measured using the proposed method.
The FID sequence consisted of an with-RF subsequence and an without-RF subsequence.
The with-RF subsequence contained a 500-microsecond sinc pulse with no gradients.
The without-RF subsequence was a 9500-microsecond delay with no RF pulses and gradients.
The efficiency was evaluated by running the proposed method
with and without Eqs. (\ref{eq:analytic_xy_derivative}) and (\ref{eq:analytic_z_derivative}).
The former and latter cases are referred to as FID-analytic and FID-update.
When these equations were not used, the without-RF subsequence was treated as
a 9500-microsecond RF pulse whose shapes were all zeros.
In either case, the with-RF subsequence was simulated with
both Eqs. (\ref{eq:update_derivatives_vector7}) and (\ref{eq:combined_transition7}).

To evaluate the proposed method in realistic condition,
acquisitions of four pulse sequences available as examples of the Pulseq\cite{Pulseq}
were simulated, and their images were reconstructed with a gridding algorithm.
Their computational times were also measured with and without combined transitions.
The sequence parameters of the pulse sequences are shown in Table \ref{tablePulseq}.
The following methods were evaluated.
\begin{description}
\item[No $T_2^{\prime}$]
A conventional method without both $T_2^{\prime}$ and combined transitions.
\item[No $T_2^{\prime}$-combined]
A conventional method without $T_2^{\prime}$ and with combined transitions.
\item[$T_2^{\prime}$]
The proposed method without combined transitions.
\item[$T_2^{\prime}$-combined]
The proposed method with combined transitions.
\end{description}
In the cases of the conventional and proposed methods,
the combined transitions were 
$\boldsymbol{C}_4(k,t_0,t_{N_{RF}})$ and
$\boldsymbol{C}_7(k,t_0,t_{N_{RF}})$, respectively.
In the case of the circles-$T_2^{*}$ phantom,
a conventional method whose $\Delta B_0$ values were sampled from a Lorentzian function with $T_2^{\prime}$
($T_2^{\prime}$-random)
was also used for simulating reconstruction images using a discrete representation.

\begin{figure}[tp]%
\centering%
\includegraphics[height=14cm]{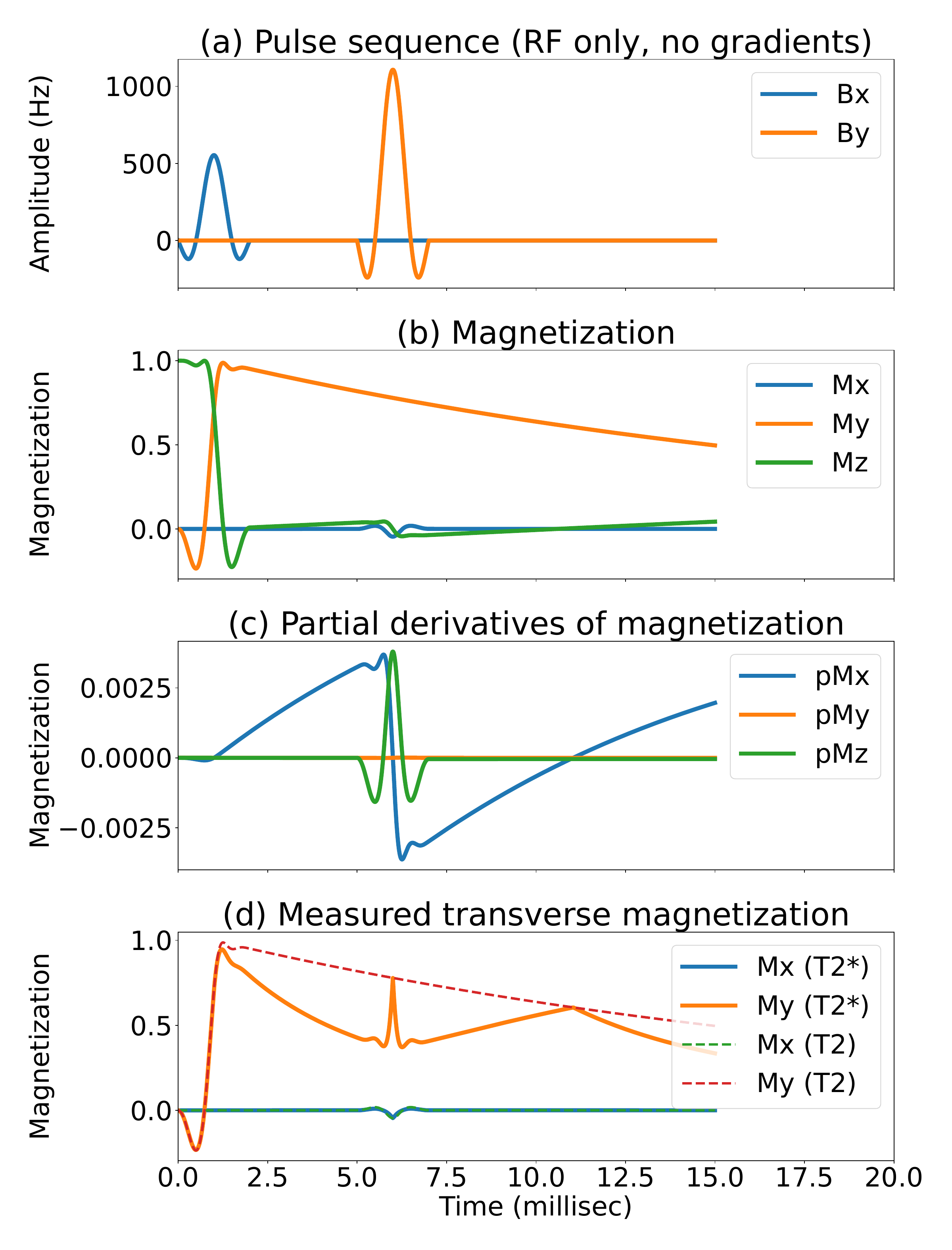}%
\caption{Temporal dynamics of (a) the simulated pulse sequence,
(b) the magnetization vector, (c) its partial derivatives,
and (d) measured transverse magnetization using an ADC with and without $T_2^{\prime}$.
As shown in (b), $M_y$ decayed monotonically with $T_2$.
As shown in (c), $\partial_{\omega} M_x$ increased monotonically
except the time around the refocusing pulse shown in (a).
By simulating (b) only, $T_2$ decay was simulated.
By simulating both (b) and (c),
$T_2^{*}$ decay could be simulated.%
}\label{figSimpleCase}
\end{figure}%
\section{Results}

\subsection{A Simple Case for Understanding the Fundamental Mechanism}

The temporal dynamics of the simulation with the proposed method
is shown in Fig. \ref{figSimpleCase}.
Whereas $M_y$ decayed monotonically with $T_2$ (Fig. \ref{figSimpleCase}(b)),
$\partial_{\omega} M_x$ increased monotonically
except the time around the refocusing pulse (Fig. \ref{figSimpleCase}(c)).
Fig. \ref{figSimpleCase}(d) represent temporal dynamics of 
the measured transverse magnetization using an ADC with and without $T_2^{\prime}$.
Whereas measured transverse magnetization using Eq. (\ref{eq:adc_simplified})
was same as $M_x$ and $M_y$ shown in Fig. \ref{figSimpleCase}(b),
measured transverse magnetization using Eq. (\ref{eq:adc_proposed})
included $T_2^{\prime}$ as expected.

\begin{table}[t]%
\caption{%
Processing times using the (a) circles-$T_2^{*}$ and
(b) brain-$T_2^{*}$ phantoms, in second.
The number of isochromats in the brain-$T_2^{*}$ phantom was
14 times greater than that in the circles-$T_2^{*}$ phantom.
With no accelerations, the computational times with $T_2^{\prime}$ was 2.0 to 2.2 times
longer than those without $T_2^{\prime}$.
The computational times of the FID-analytic was 19 times faster than those of the FID-update.
With combined transitions, the computational times were up to up to 17 times faster if many RF pulses were repeated.%
}\label{tableProctime}
\centering%
\begin{tabular}{|l|r|r|r|r|r|r|} \hline
\multicolumn{7}{|l|}{ (a) Circles-$T_2^{*}$ (20068128 isochromats) } \\ \hline
 &FID-update&FID-analytic&SPGR&RARE&EPI&Spiral \\ \hline
No $T_2^{\prime}$ & 18.36 & 0.97 & 750.11 & 614.53 & 34.52 & 153.92\\ \hline
$T_2^{\prime}$ & 39.93 & 2.14 & 1609.19 & 1334.54 & 72.65 & 315.48\\ \hline
No $T_2^{\prime}$ + combined & 34.26 & 2.00 & 45.51 & 52.65 & 50.67 & 143.20\\ \hline
$T_2^{\prime}$ + combined & 86.99 & 5.16 & 93.37 & 111.27 & 117.16 & 321.44\\ \hline
\multicolumn{7}{|l|}{ (b) Brain-$T_2^{*}$ (280152000 isochromats) } \\ \hline
 &FID-update&FID-analytic&SPGR&RARE&EPI&Spiral \\ \hline
No $T_2^{\prime}$ & 251.41 & 13.18 & 10159.93 & 8443.54 & 483.89 & 2077.85\\ \hline
$T_2^{\prime}$ & 535.39 & 28.18 & 22020.83 & 17843.24 & 974.39 & 4415.70\\ \hline
No $T_2^{\prime}$ + combined & 462.12 & 27.45 & 611.37 & 710.08 & 677.45 & 1924.95\\ \hline
$T_2^{\prime}$ + combined & 1229.51 & 73.59 & 1323.49 & 1563.82 & 1654.42 & 4511.43\\ \hline
\end{tabular}%
\end{table}%
\begin{figure}[tp]%
\centering%
\includegraphics[width=\linewidth]{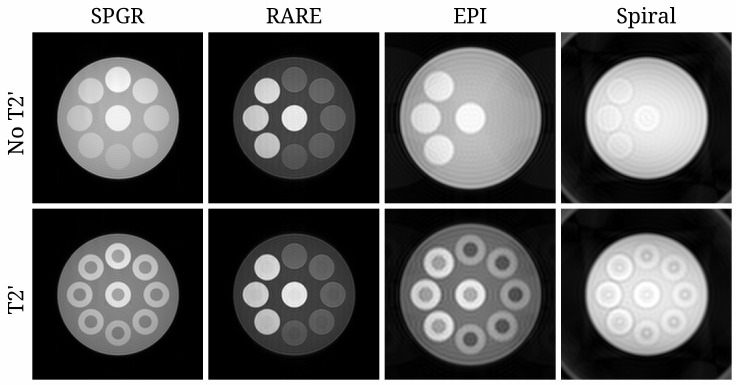}%
\caption{Reconstructed images using the circles-$T_2^{*}$ phantom with four pulse sequences.
Whereas the results without $T_2^{\prime}$ did not simulate inner concentric cylinders,
the results with $T_2^{\prime}$ simulated inner concentric cylinders whose $T_2^{*}$ values were
different from outer concentric cylinders in the cases of gradient-echo sequences.
It can also be confirmed that $T_2$ contrast images were reconstructed in the cases of the RARE sequence.
}\label{figCirclesResults}
\end{figure}%
\begin{figure}[tp]%
\centering%
\includegraphics[width=\linewidth]{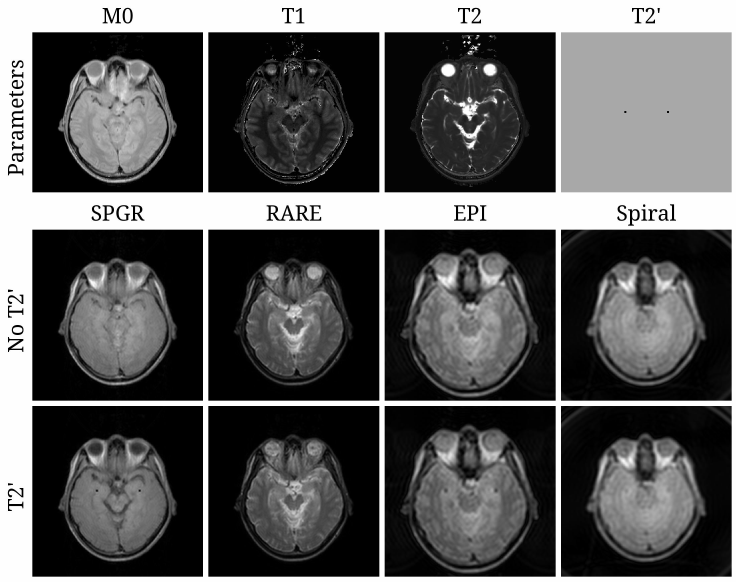}%
\caption{Reconstructed images using the brain-$T_2^{*}$ phantom with four pulse sequences.
As shown in the $T_2^{\prime}$ map, there were two dots whose $T_2^{\prime}$ values were low.
By comparing the results with and without $T_2^{\prime}$,
two dots were reconstructed in the cases of gradient-echo sequences only if $T_2^{\prime}$ was simulated.
}\label{figBrain}
\end{figure}%
\subsection{Evaluations with Multiple Isochromats}

The computational times of both FID and Pulseq sequences are shown in Table \ref{tableProctime}.
In all the evaluated cases,
both number of isochromats and processing times of the brain-$T_2^{*}$ phantom were
approximate 14 times greater than those of the circles-$T_2^{*}$ phantom.
The computational times of the FID-analytic was 19 times faster than those of the FID-update.
In the cases of the simulations with and without combined transitions,
the computational times with $T_2^{\prime}$ was 2.0 to 2.2 times and 2.0 to 2.7 times
longer than those without $T_2^{\prime}$, respectively.

Reconstructed images using the circles-$T_2^{*}$ phantom
are shown in Fig. \ref{figCirclesResults}.
Whereas the results without $T_2^{\prime}$ did not simulate inner concentric cylinders,
the results with $T_2^{\prime}$ simulated inner concentric cylinders whose $T_2^{*}$ values were
different from outer concentric cylinders in the cases of gradient-echo sequences.
It was also confirmed that
only the outer concentric cylinders were reconstructed in the cases of the RARE sequence.

Reconstructed images using the brain-$T_2^{*}$ phantom are shown in Fig. \ref{figBrain}.
In the cases of the circles-$T_2^{*}$ phantom,
As shown in the $T_2^{\prime}$ map, there were two dots whose $T_2^{\prime}$ values were low.
By comparing the results with and without $T_2^{\prime}$,
two dots were reconstructed in the cases of gradient-echo sequences only if $T_2^{\prime}$ was simulated.

\section{Discussion}

The main contributions of this paper are
to propose the novel simulation method of $T_2^{\prime}$ 
without relying on 100+ isochromats,
and to demonstrate their efficiency of the proposed method experimentally.
As derived in the Theory section,
$T_2^{\prime}$ can be modeled by simulating $\boldsymbol{M}_7(k,t)$ instead of $\boldsymbol{M}_4(k,t)$
and using a linear phase model.
Whereas the computational times for simulating $T_2^{\prime}$ were
2.0-2.7 times longer, the preciseness of the $T_2^{\prime}$ decay was theoretically better than
simulations of $T_2^{\prime}$ with 100+ isochromats
using conventional methods as shown in Fig. \ref{figLorentizian}.

The simple case demonstrated that
only one isochromat were required for $T_2^{\prime}$ simulations
if the partial derivatives of the magnetizations were available.
Without this information, only $T_2$ decays could be simulated as shown in Fig. \ref{figSimpleCase} (d).
In addition, the temporal dynamics of the partial derivatives gave an insight into
the relative strength of the $T_2^{\prime}$ effects.
It was not easy to get such information from the magnetizations only even if 100+ isochromats were used.
Therefore, this simulation method is expected to be useful for understanding the principle of $T_2$ and $T_2^{*}$.

The evaluations with both the circles-$T_2^{*}$ and brain-$T_2^{*}$ phantoms showed that
practical $T_2^{\prime}$ simulations were possible
without increasing the number of isochromats for modeling a Lorentzian function.
The differences of the $T_2^{*}$ values were easily confirmed as the visibilities of
the inner concentric cylinders of the circles-$T_2^{*}$, and the dots of the brain-$T_2^{*}$ phantoms.
In addition, the computational times showed that
the combined transitions\cite{Taniguchi1,Taniguchi2,Scholand,Takeshima}
were still efficient even when the size of the magnetization vectors were increased from 4 to 7.
These results showed that
combined transitions were useful in the cases of RF pulses which were reused at least 3 times.
RF pulses which do not meet this criterion should be simulated directly
with Eq. (\ref{eq:update_derivatives_vector7}) whenever possible.

A fundamental limitation of the proposed method is utilization of the linear-phase model in entire Lorentzian functions.
This model is incorrect if RF pulses used in a pulse sequence are highly frequency dependent.
Future work includes simulations of such RF pulses more precisely.

\section{Conclusion}

Both theory and results showed that
the proposed methods simulated $T_2^{\prime}$ efficiently by utilizing
a linear model with a Lorentzian function,
analytic solutions, and combined transitions.
Their fundamental mechanism was experimentally confirmed using a simple case, and
their effectiveness was demonstrated using a house-made MR simulator with 284 million isochromats.

\section{Disclosure Statement}

Hidenori Takeshima is an employee of Canon Medical Systems Corporation.

\appendix
\section{Matrix Representations of Transitions and Their Partial Derivatives}\label{appendix:transition}

According to the implementations used an asymmetric operator splitting \linebreak
method \cite{Graf2},
the transition can be approximated as $\boldsymbol{D}_4(k, \Delta t) \boldsymbol{R}_4(k, t, t+\Delta t)$ where
the matrices $\boldsymbol{D}_4$ and $\boldsymbol{R}_4$ represent relaxation and rotation, respectively.
These matrices are given as
\begin{equation}\label{eq:matrix_D4}
\boldsymbol{D}_4 =
\left(
\begin{array}{cc}
  \mbox{\Large $\boldsymbol{D}_3$} &
    \begin{matrix}
      0 \\
      0 \\
      1-\exp(-\Delta t \text{⁄} T_1)
    \end{matrix} \\
  \begin{matrix}
  0 & 0 & 0
  \end{matrix}
   & 1
\end{array}
\right)
\text{,}
\end{equation}
\begin{equation}\label{eq:matrix_D3}
\boldsymbol{D}_3 =
\begin{pmatrix}
\exp(-\Delta t \text{⁄} T_2) & 0 & 0 \\
0 & \exp(-\Delta t \text{⁄} T_2) & 0 \\
0 & 0 & \exp(-\Delta t \text{⁄} T_1)
\end{pmatrix}
\text{,}
\end{equation}
\begin{equation}\label{eq:matrix_R4}
\boldsymbol{R}_4=
\left(
\begin{array}{cc}
  \mbox{\Large $\boldsymbol{R}_3$} &
    \begin{matrix}
      0 \\
      0 \\
      0
    \end{matrix} \\
  \begin{matrix}
  0 & 0 & 0
  \end{matrix}
  & 1
\end{array}
\right)
\text{, and}
\end{equation}
\begin{equation}\label{eq:matrix_R3}
\boldsymbol{R}_3=
\begin{pmatrix}
      \cos \theta + u_x u_x(1 - \cos \theta) &
- u_z \sin \theta + u_x u_y(1 - \cos \theta) &
  u_y \sin \theta + u_x u_z(1 - \cos \theta) \\
  u_z \sin \theta + u_x u_y(1 - \cos \theta) &
      \cos \theta + u_y u_y (1 - \cos \theta) &
- u_x \sin \theta + u_y u_z (1 - \cos \theta) \\
- u_y \sin \theta + u_x u_z (1 - \cos \theta) &
  u_x \sin \theta + u_y u_z (1 - \cos \theta) &
      \cos \theta + u_z u_z (1 - \cos \theta) 
\end{pmatrix}
\end{equation}
where $(u_x,u_y,u_z) = \boldsymbol{B}(k,t) \text{⁄} |\boldsymbol{B}(k,t)|$ and
$ \theta = -\gamma |\boldsymbol{B}(k,t)|\Delta t$.

As used in Jochimsen et al.\cite{Jochimsen},
the rotation matrix $\boldsymbol{R}_3(k, t, t+\Delta t)$ can be further approximated by
processing rotations in the $z$ direction separately
as $\boldsymbol{R}_3(k, t, t+\Delta t) = \boldsymbol{R}_{3,xy}(k, t, t+\Delta t) \boldsymbol{R}_{3,z}(k, t, t+\Delta t)$.
The matrix $\boldsymbol{R}_{3,xy}(k, t, t+\Delta t)$ corresponds to
the matrix $\boldsymbol{R}_3(k,t)$ whose $B_z(k,t)$ of $\boldsymbol{B}(k,t)$ is replaced by zero.
The matrix $\boldsymbol{R}_{3,z}(k, t, t+\Delta t)$ corresponds to
the matrix $\boldsymbol{R}_3(k,t)$ whose $B_x(k,t)$ and $B_y(k,t)$ of $\boldsymbol{B}(k,t)$ are replaced by zeros.
For the sake of convenience,
$\boldsymbol{R}_{3,xy}(k, t, t+\Delta t)$ and
$\boldsymbol{R}_{3,z}(k, t, t+\Delta t)$ are defined similarly.
The matrix representations of these matrices are
\begin{equation}\label{eq:matrix_Rxy}
\boldsymbol{R}_{3,xy}=
\begin{pmatrix}
      \cos \theta_{xy} + u_x u_x(1 - \cos \theta_{xy}) &
                    u_x u_y(1 - \cos \theta_{xy}) &
  u_y \sin \theta_{xy} \\
                    u_x u_y(1 - \cos \theta_{xy}) &
      \cos \theta_{xy} + u_y u_y (1 - \cos \theta_{xy}) &
- u_x \sin \theta_{xy} \\
- u_y \sin \theta_{xy} &
  u_x \sin \theta_{xy} &
      \cos \theta_{xy} 
\end{pmatrix}
\text{, and}
\end{equation}
\begin{equation}\label{eq:matrix_Rz}
\boldsymbol{R}_{3,z}=
\begin{pmatrix}
      \cos \theta_z &
    - \sin \theta_z &
  0 \\
      \sin \theta_z &
      \cos \theta_z &
  0 \\
0 & 0 & 1 \\
\end{pmatrix}%
\end{equation}
where $\theta_{xy}$ and $\theta_z$ represent $\theta$ calculated for 
$\boldsymbol{R}_{3,xy}(k, t, t+\Delta t)$ and $\boldsymbol{R}_{3,z}(k, t, t+\Delta t)$, respectively.

The partial derivative of the matrix $\boldsymbol{R}_{3,z}(k, t, t+\Delta t)$
with respect to $p$ is
\begin{equation}\label{eq:matrix_Rz_partial3}
\partial p \boldsymbol{R}_{3,z}
=
\frac{\partial \theta_z}{\partial p}
\begin{pmatrix}
    - \sin \theta_z &
    - \cos \theta_z &
  0  \\
      \cos \theta_z &
    - \sin \theta_z &
  0  \\
0 & 0 & 0
\end{pmatrix}%
\end{equation}
where $\partial \theta_z/\partial p = - \Delta t$ in the case of $p = \omega$.

The $7 \times 7$ matrices $\boldsymbol{D}_7(k, \Delta t)$
and
$\boldsymbol{R}_7(k, \Delta t)$,
introduced in Eq. (\ref{eq:update_derivatives_vector7}), are given as
\begin{equation}\label{eq:matrix_D7}
\boldsymbol{D}_7
=
\left(
\begin{array}{cc}
  \boldsymbol{D}_4 & \boldsymbol{0} \\
  \boldsymbol{0} & \boldsymbol{D}_3
\end{array}
\right)
\text{, and}
\end{equation}
\begin{equation}\label{eq:matrix_R7}
\boldsymbol{R}_7
=
\left(
\begin{array}{cc}
  \boldsymbol{R}_4 & \boldsymbol{0} \\
  \boldsymbol{R}_{3,xy} \partial_p \boldsymbol{R}_{3,z} &
  \boldsymbol{R}_{3,xy} \boldsymbol{R}_{3,z}
\end{array}
\right)
\text{.}
\end{equation}

\bibliographystyle{elsarticle-num}
\bibliography{paperref}

\newpage
{\Large\centering%
Supplementary Materials}

\section*{S1. Pseudo-code of the Key Components in the Proposed Method}

The key components in the proposed method are
simulations of $T_2^{\prime}$, 
fast simulations with analytic solutions, 
and
fast simulations with combined transitions. 
For better understanding of the proposed method,
this supplementary document provides pseudo-code for these key components.

The first key component is a sampling method for simulating $T_2^{\prime}$.
This method includes simluations of Eqs. (\ref{eq:adc_proposed}) and (\ref{eq:update_derivatives_vector7}).
Assuming that $\boldsymbol{M}_7(k,t)$'s are simulated at all $t$'s to be sampled via ADC.
Simplified pseudo-code for sampling data can be written as follows.
In the following pseudo-code, receiver coil sensitivity is omitted.

\begin{algorithmic}

\Function{M7\_sampling\_T2star}{$M_{7,t}$,$M_0$,$T_2^{\prime}$}
    \For{$t$ in all times}
        \State M7sampling $\gets$ $M_{7,t}$[t] \Comment{M7 at the time t}
        \State $M_x$ $\gets$ M7sampling[0] \Comment{Extract elements of M7}
        \State $M_y$ $\gets$ M7sampling[1]
        \State $\partial M_x$ $\gets$ M7sampling[4]
        \State $\partial M_y$ $\gets$ M7sampling[5]
        \State tmp $\gets$ $\sqrt{M_x^2 + M_y^2}$ \Comment{Compute a partial derivative of phi}
        \State tmp $\gets$ max(tmp, $1^{-10}$) \Comment{Avoid zero-division errors}
        \State $\partial \phi$ $\gets$ ($M_x*\partial M_y - M_y*\partial M_x$) / tmp
        \Comment{Compute a sample with T2-prime decay at t}
        \State sample $\gets$ $M_0 * (M_x + 1j*M_y) * \exp(-(1/T_2^{\prime}) * |\partial \phi|$)
        \State allsamples[t] $\gets$ sample
    \EndFor
    \State \Return allsamples
\EndFunction

\end{algorithmic}

The second key component is a fast simulation method using analytic solutions.
The analytic solutions of the partial derivative elements of $\boldsymbol{M}_7(k,t)$ are derived as
Eqs. (\ref{eq:analytic_xy_derivative}) and (\ref{eq:analytic_z_derivative}).
To compute remaining elements of $\boldsymbol{M}_7(k,t)$,
Eqs. (\ref{eq:analytic_xy}) and (\ref{eq:analytic_z}) are also required.
Pseudo-code for updating with these analytic solutions can be written as follows.

\begin{algorithmic}

\Function{analytic\_M7\_update}{$M_7$, integral\_Bz, duration, $T_1$, $T_2$}
    \State $M_{xy}$ $\gets$ $M_7$[0] + 1j*$M_7$[1] \Comment{Extract elements of M7}
    \State $M_z$ $\gets$ $M_7$[2]
    \State $\partial M_{xy}$ $\gets$ $M_7$[4] + 1j*$M_7$[5]
    \State $\partial M_z$ $\gets$ $M_7$[6]
    \State tmp $\gets$ exp(-1j * integral\_Bz) \Comment{Precompute a rotation}
    \State $\partial M_{xy}$ $\gets$ $\exp(-\text{duration}/T_2) * (\partial M_{xy} * tmp - 1j * \text{duration} * M_{xy} * tmp)$
    \State \Comment{Compute partial derivatives of Mxy and Mz}
    \State $\partial M_z$ $\gets$ $\partial M_z * \exp(-\text{duration}/T_1)$
    \State $M_{xy}$ $\gets$ $M_{xy} * \exp(-\text{duration}/T_2) * tmp$ \Comment{Compute Mxy and Mz}
    \State $M_z$ $\gets$ $1 + (M_z - 1) * \exp(-\text{duration}/T_1)$
    \State \Return [$M_{xy}$.real, $M_{xy}$.imag, $M_z$, 1, $\partial M_{xy}$.real, $\partial M_{xy}$.imag, $\partial M_z$] \Comment{Return updated M7}
\EndFunction

\end{algorithmic}

The third key component is a fast simulation method using combined transitions.
Straightforward computation of $\boldsymbol{C}_7(k,t_0,t_{N_{RF}})$ is possible but increases the computational times.
To avoid $7 \times 7$ matrix multiplications in computing $\boldsymbol{C}_7(k,t_0,t_{N_{RF}})$,
simulations of 7 linearly independent vectors of $\boldsymbol{M}_7(k,t)$ with Eq. (\ref{eq:update}) can be used instead.
By using a set 7 unit vectors whose fourth elements are set to 1,
pseudo-code for computing $\boldsymbol{C}_7(k,t_0,t_{N_{RF}})$ without $7 \times 7$ matrix multiplications 
can be written as follows.

\begin{algorithmic}

\Function{matrix\_C7\_combined}{$B_x, B_y, B_z, \Delta t, T_1, T_2$}
    \State \Comment{Initial vectors. The fourth elements are constants and thus are set to 1.}
    \State M7 = [
    \State       [1,0,0,1,0,0,0],
    \State       [0,1,0,1,0,0,0],
    \State       [0,0,1,1,0,0,0],
    \State       [0,0,0,1,0,0,0],
    \State       [0,0,0,1,1,0,0],
    \State       [0,0,0,1,0,1,0],
    \State       [0,0,0,1,0,0,1]
    \State      ]
    \For{$t$ in all times}
        \For{v in $4,5,6$}
            \Comment{Update partial derivatives only on updating the last three vectors.}
            \State M7[v][4:7] $\gets$ update\_pM3($\text{M7[v]}, B_x[t], B_y[t], B_z[t], \Delta t, T_1, T_2$)
        \EndFor
        \For {v in $0,1,2,3$}
            \Comment{Update all elements.}
            \State M7[v] $\gets$ update\_M7($\text{M7[v]}, B_x[t], B_y[t], B_z[t], \Delta t, T_1, T_2$)
        \EndFor
        \For{v in $4,5,6$}
            \Comment{Update M4 components of the last three vectors.}
            \State M7[v][0:4] $\gets$ M7[3][0:4]
        \EndFor
    \EndFor
    \State \Comment{Since the fourth elements of the initial vectors are set to 1,}
    \For {v in $0,1,2,4,5,6$} \Comment{the following compensations are required.}
        \State M7[v] $\gets$ M7[v] - M7[3]
    \EndFor
    \State \Return transpose\_matrix(M7)
\EndFunction

\end{algorithmic}

It is worth noting that update of $\boldsymbol{M}_4(k,t)$ can be omitted on updating the last three vectors
since $\boldsymbol{M}_4(k,t)$ is independent from $\partial_{\omega} \boldsymbol{M}_4(k,t)$.

\section*{S2. Parameter Mapping Method for Generating the Brain-$T_2^{*}$ Phantom}

Four T1-weighted (T1W) images, and four T2-weighted (T2W) images were acquired for generating parameter maps.
All acquisitions shared the matrix and field-of-view (FOV) sizes. 
The following table describes pulse sequences used in the volunteer scan.
\begin{table}[h]%
\centering%
\begin{tabular}{|l|l|l|} \hline
 & T1W (4 acquisitions) & T2W \\ \hline
Sequence type & 2D SE & 2D RARE with 4 echos \\ \hline
Reconstruction matrix & $512 \times 480 \times 48$ & $512 \times 480 \times 48$ \\ \hline
FOV & $256 \times 240 \times 240 \,\text{mm}^3$ & $256 \times 240 \times 240 \,\text{mm}^3$ \\ \hline
Num. echos & 1 & 4 \\ \hline
TE & 10 ms & 20, 60, 100, 140 ms \\ \hline
TR & 500, 800, 1100, 1500 ms & 4500 ms \\ \hline
\end{tabular}
\end{table}

The $T_1$ and $T_2$ values were estimated using the following parameter mapping method.
For each pixel,
the best $T_1$ value was searched from 200 logarithmically-spaced candidate values.
The candidate values were put in the range of 10 to 4500 milliseconds.
Similarly, for each pixel,
the best $T_2$ value was searched from 200 logarithmically-spaced candidate values.
The candidate values were put in the range of 10 to 2500 milliseconds.

After the $T_2$ values were estimated,
the $M_0$ values were estimated using all T2W images with a least-squares method.
In the estimation of the $M_0$ values, the $T_1$ values were ignored.
The pixels whose $M_0$ values were lower than a threshold were removed.
To remove undesirable noises,
pixels whose $T_1$ values were the maximum value (4500 milliseconds) were also removed.

\end{document}